\documentclass[twocolumn,10pt,showpacs,amsmath,lang]{revtex4}
\usepackage[cp866]{inputenc}
\usepackage[english]{babel}
\usepackage{graphicx}
\usepackage{bm}
\usepackage{epsf}
\usepackage{amssymb,amsmath}

\begin{document}

\title
{Electron thermal conductivity owing to collisions
between degenerate electrons}

\author{P.~S.~Shternin 
        and D.~G.~Yakovlev}
\affiliation{
        Ioffe Physical Technical Institute,
        Politekhnicheskaya 26, 194021 Saint-Petersburg, Russia}
\date{\today}

\begin{abstract}
We calculate the thermal conductivity of electrons
produced by electron-electron Coulomb scattering in a
strongly degenerate electron gas taking into account the Landau damping of
transverse plasmons. The Landau damping strongly reduces
this conductivity in the domain of ultrarelativistic electrons
at temperatures below the electron plasma temperature.
In the inner crust of a neutron star at temperatures
$T \lesssim 10^7$~K this thermal conductivity
completely dominates over the electron conductivity due to electron-ion
(electron-phonon)
scattering and becomes competitive with the the electron conductivity
due to scattering of electrons by impurity ions.
\end{abstract}


\maketitle

\section{Introduction}
\label{introduc}

The electron thermal conductivity is an important
kinetic property of plasmas in experimental devices,
in metals and semiconductors and in astrophysical objects.
It has been studied for a long time and described
in textbooks (see, e.g., Ziman \cite{ziman60},
Lifshitz and Pitaevski{\u\i}
\cite{lp81}). We will show that some aspects of
this problem have to be revised.

Specifically, we will consider the thermal
conductivity of degenerate electrons. It is important
in metals, semiconductors as well as
in degenerate cores of evolved stars (giants,
supergiants and, most importantly, white dwarfs),
and in envelopes of neutron stars.
It is needed to study cooling of white dwarfs (e.g.,
Prada Moroni and Straniero \cite{ps02})
and nuclear explosions of massive
white dwarfs as type Ia supernovae
(e.g.,  Baraffe, Heger and Woosley \cite{bhw04}).
In neutron star envelopes it is required for many reasons ---
for simulating cooling of
isolated neutron stars (see, e.g., Lattimer et al.\ \cite{lvpp94},
Gnedin, Yakovlev and Potekhin \cite{gyp01},
Page, Geppert and Weber \cite{pgw06});
for studying heat propagation and
thermal relaxation in a neutron star envelope in
response to the various
processes of crustal energy release. In particular,
these processes include pulsar glitches
(e.g., Larson and Link \cite{ll02}, and references therein),
deep crustal heating of accreting neutron stars
in soft X-ray transients
(e.g., Ushomirsky and Rutledge \cite{ur01}), superbursts as
powerful nuclear explosions at the base of the outer crust
of an accreting neutron star
(e.g., Strohmayer and Bildsten \cite{sb06};
Page and Cumming \cite{pc05}). 

It is well known,
that the electrons give the main contribution
into the thermal conductivity of strongly degenerate
matter. Their thermal conductivity
can be written as
\begin{equation}
     \kappa_e={\pi^2 T k_{\rm B}^2 n_e \over 3m_e^*\nu_e},
     \quad \nu_e=\nu_{ei}+\nu_{ee}.
\label{kappa}
\end{equation}
Here, $T$ is the temperature, $k_{\rm B}$ is the Boltzmann constant,
$n_e$ is the number density of electrons, $m_e^*=\mu/c^2$,
$\mu$ is the electron chemical potential
(including the rest-mass term), and $\nu_e$ is
the total effective electron collision frequency.
The latter frequency is the sum of the partial collision frequencies
$\nu_{ei}$ and $\nu_{ee}$. In $\nu_{ei}$ we include
all collisions of electrons mediated by
their interactions with ions (direct Coulomb scattering
of electrons by ions in an ion gas or liquid;
electron-phonon scattering in an ion crystal
and electron scattering by impurity ions at low
temperatures).  Evidently, Eq.\ (\ref{kappa})
can be rewritten as
\begin{eqnarray}
 && { 1 \over \kappa_e} ={ 1 \over \kappa_{ei}} +{ 1 \over \kappa_{ee}},
  \quad
   \kappa_{ei}={\pi^2 T k_{\rm B}^2 n_e \over 3m_e^*\nu_{ei}},
\nonumber \\
 &&  \kappa_{ee}={\pi^2 T k_{\rm B}^2 n_e \over 3m_e^*\nu_{ee}},
\label{kappae+i}
\end{eqnarray}
where the partial conductivities
$\kappa_{ei}$ and $\kappa_{ee}$
are determined, respectively, by the
$ei$ and $ee$ collisions.

It is widely believed that the dominant contribution
into $\kappa_e$ comes from the $ei$ collisions.
The associated partial conductivity $\kappa_{ei}$
has been studied in a number of papers
(e.g., Potekhin et al.\ \cite{pbhy99}, Gnedin et al.\ \cite{gyp01}
and references therein). The partial thermal
conductivity $\kappa_{ee}$ owing to the $ee$ collisions
was calculated by Lampe \cite{lampe68}, Flowers and Itoh \cite{fi76},
Urpin and Yakovlev \cite{uy80}, and Timmes \cite{timmes92}.
These calculations were summarized by Potekhin et al.\ \cite{pbhy99}.
The main result was that
the $ee$ collisions are mainly negligible,
except for a hot low-density plasma of light
elements (from hydrogen to carbon) with the temperature
a few times lower than the electron degeneracy temperature
(in that case the $ee$ collisions did affect $\kappa_{e}$
but never dominated
as long as the electrons were degenerate; see
Lampe \cite{lampe68}; Urpin and Yakovlev \cite{uy80}).

However, all these calculations have
neglected an important effect of the Landau damping of
the $ee$ interaction
owing to the exchange of transverse plasmons. In the context of
transport properties of dense matter
this effect was studied by Heiselberg and Pethick \cite{hp93}
for degenerate quark plasmas. Similar effects have
recently been analyzed by Jaikumar, Gale and Page \cite{jgp05}
for neutrino bremsstrahlung radiation in the $ee$
collisions. Here we reconsider $\kappa_{ee}$
including the effects of the Landau damping and show
that $\kappa_{ee}$ is actually much more
important than thought before.

\section{Formalism}
\label{formalism}

We analyze the thermal conductivity $\kappa_{ee}$
of strongly degenerate electrons under the conditions
that the electrons constitute an almost ideal
and uniform Fermi gas and collide between
themselves and with plasma ions. Although the results will be
more important for ultrarelativistic electrons,
our consideration will be general and valid for both,
relativistic and non-relativistic, cases.
We will use the same standard variational
approach with the simplest trial non-equilibrium electron
distribution function as Flowers and Itoh \cite{fi76}.
The calculations are similar to those
performed by Heiselberg and Pethick \cite{hp93}
for the thermal conductivity
of quarks; thus we omit the details.

Following Heiselberg and Pethick \cite{hp93} we have
\begin{eqnarray}
   \nu_{ee} & = & {3  \hbar^3 \over 8 v_e p_e^2 (k_{\rm B}T)^3 } \, \int
   {{\rm d}\bm{p}_1\, {\rm d}\bm{p}_2\,{\rm d}\bm{p}'_1\,{\rm d}\bm{p}'_2
   \over (2 \pi \hbar)^{12}} 
\nonumber \\   
  && \times  W(12|1'2')\,f_1 f_2 (1-f'_1)(1-f'_2)
\nonumber \\
   & & \times \left[
   \bm{v}_1 \, (\varepsilon_1 - \mu) +
   \bm{v}_2 \, (\varepsilon_2 - \mu) -
   \bm{v}'_1 \, (\varepsilon'_1 -\mu) \right.
\nonumber \\   
 &&  - \left.
   \bm{v}'_2 \, (\varepsilon'_2 -\mu)
   \right]^2, 
\label{nuee}
\end{eqnarray}
where $p_e = \hbar\, (3 \pi^2 n_e)^{1/3}$ and $v_e=p_e/m_e^*$ are,
respectively, the electron Fermi momentum and Fermi velocity;
the integration is over all allowable $ee$ collisions
$\bm{p}_1 \bm{p}_2 \to \bm{p}'_1 \bm{p}'_2$; $\bm{p}$ is
an electron momentum, $\bm{v}$ its velocity,
$\varepsilon$ its energy; primes refer to electrons after
a collision event; $f$ is the electron Fermi-Dirac distribution,
and $W(12|1'2')$ is the differential transition
probability
\begin{eqnarray}
     W(12|1'2')&=&4\,(2\pi\hbar)^6\,
     \delta(\varepsilon'_1+\varepsilon'_2-\varepsilon_1-\varepsilon_2)\,
\nonumber \\     
     && \times 
     \delta(\bm{p}'_1+\bm{p}'_2-\bm{p}_1-\bm{p}_2)\,|M_{fi}|^2,
\label{W}
\end{eqnarray}
$|M_{fi}|^2$ being the squared matrix element summed over electron
spin states. The symmetry factors required to
avoid double counting of the same initial and final electron states
are also included into $|M_{fi}|^2$.

The multi-dimensional integral (\ref{nuee})
is simplified further using the standard
angular-energy decomposition
(separating integrations over particle energies
and orientations of their momenta; e.g., Shapiro
and Teukolsky \cite{st83}).
Because the electrons are
strongly degenerate, they can participate in thermal conduction
only if their energies are close to the Fermi level.
Accordingly, their momenta can be placed at
the Fermi surface in angular integrals whenever possible.
Characteristic energy transfers $\hbar \omega \equiv
\varepsilon'_1- \varepsilon_1$ in collisions
of strongly degenerate particles are small,
$\hbar \omega \lesssim k_{\rm B} T$.
Momentum transfers
$\hbar \bm{q} \equiv \bm{p}'_1-\bm{p}_1$ are also small,
$\hbar q \ll p_e$, owing to a long-range nature of the Coulomb interaction.
We will use this small-momentum-transfer approximation
throughout the paper.
Typical values of $\hbar q$ are determined
by plasma screening of the Coulomb interaction.

The plasma screening
was thoroughly analyzed by Heiselberg and Pethick \cite{hp93}.
These authors studied quark-quark interaction through
one-gluon exchange in the weak-coupling limit which
is very similar to the Coulomb interaction in an ordinary plasma.
The matrix element for an $ee$ scattering event is
$M_{fi}=M^{(1)}_{fi}+
M^{(2)}_{fi}$, where $M^{(1)}_{fi}$ and $M^{(2)}_{fi}$\
correspond to the channels  $1 \to 1';2 \to 2'$
and $1 \to 2';2 \to 1'$, respectively.
For instance,
\begin{equation}
      M^{(1)}_{fi} \propto \frac{J_{1'1}^{(0)}J_{2'2}^{(0)}}{q^2+\Pi_l}-
      \frac{\bm{J}_{t1'1} \bm{\cdot} \bm{J}_{t2'2}}{q^2-\omega^2/c^2+\Pi_t} ,
\label{matelement}
\end{equation}
where $J_{e'e}^{(\nu)}=(J_{e'e}^{(0)},\bm{J}_{e'e})=
c\,(\bar{u}_{e'}\gamma^\nu u_e)$ is the transition 4-current
($\nu$=0, 1, 2, 3),
$\bm{J}_{te'e}$ is the component of $\bm{J}_{e'e}$ transverse to $\bm{q}$,
$\gamma^\nu$ is a Dirac matrix,
$u_e$ is a normalized electron bispinor ($\bar{u}_e u_e=2 m_ec^2$),
and $\bar{u}_e$ is a Dirac conjugate (see, e.g.,
Berestetski{\u\i}, Lifshitz  and Pitaevskii \cite{blp82}).
The longitudinal component of $\bm{J}_{e'e}$
(parallel to $\bm{q}$) is related
to the time-like (charge density) component $J_{e'e}^{(0)}$
via current continuity equation; it is excluded   from
Eq.~(\ref{matelement}), see Ref.\ \cite{hp93}.

The polarization functions $\Pi_l$ and $\Pi_t$ in Eq.\ (\ref{matelement})
depend on $\omega$ and $q$ and
describe the plasma screening of
the $ee$ interaction through the exchange of longitudinal
and transverse plasmons, respectively. In the classical
limit ($\hbar q \ll p_e$ and $\hbar \omega \ll v_e p_e$),
taking into account the polarization produced by degenerate electrons
in the first-order random phase approximation,
one has (e.g., Alexandrov, Bogdankevich and Rukhadze \cite{abr84})
\begin{equation}
    \Pi_l = q_0^2 \, \chi_l(x), \qquad
    \Pi_t = (q_0 v_e/c)^2 \, \chi_t(x),
\label{polariz}
\end{equation}
where $x=\omega /(q v_e)$,
\begin{eqnarray}
     \chi_l(x) &=&  1- {x \over 2}\,
    \ln \left( x+1 \over x-1 \right) ,
\nonumber \\
    \chi_t(x) &=&  {x^2 \over 2} +
    { x(1-x^2) \over 4}\, \ln \left( x+1 \over x-1 \right) ,
\label{chi}
\end{eqnarray}
with
\begin{equation}
    \hbar^2 q_0^2={ 4 e^2 p_e^2 /( \pi \hbar v_e) },
\label{qscreen}
\end{equation}
$q_0$ being the ordinary Thomas-Fermi electron
screening wavenumber. Particularly, in the limit of
$\omega \to 0$ and $\omega/q \ll v_e$ we have
\begin{equation}
   \chi_l = 1, \qquad
   \chi_t= i \,
   \pi \omega /(4q v_e).
\label{polariz1}
\end{equation}

According to Eqs.\ (\ref{matelement}) and (\ref{polariz}),
the plasma screening of the $ee$ current interaction
via the exchange of transverse plasmons
(the second term in Eq.\ (\ref{matelement})) is different
from the screening of the charge interaction
via the exchange of
longitudinal plasmons (the first term). The difference results from
the difference of the polarization functions
$\Pi_t$ and $\Pi_l$ and has been neglected
in all previous calculations of the electron thermal
conductivity (where one has commonly set $\Pi_t=\Pi_l=q_0^2$,
and $q^2-(\omega/c)^2+\Pi_t=q^2+q_0^2$).
Naturally, this difference is expected to be small for
non-relativistic electrons ($v_e \ll c$), where
the transverse current interaction term
is small in the matrix element. We will see that
the difference becomes important for relativistic
electrons ($v_e \approx c$).

Let us recall that
the functions $\Pi_l$ and $\Pi_t$ have real parts
which describe plasmon refraction, and imaginary parts
which describe plasmon absorption. In the $ee$ scattering
we deal with low-energy virtual plasmons,
$\hbar \omega \lesssim k_{\rm B}T$.
As seen from Eq.~(\ref{polariz1}), longitudinal
plasmons undergo refraction which results in the
Debye-type (Thomas-Fermi) screening of the Coulomb
interaction, with the screening momentum
(inverse screening length) $q_0$.
As for transverse plasmons, they mainly undergo
collisionless absorption (that is the Landau
damping) by degenerate electrons. Their effect is
drastically different from the effect of longitudinal
plasmons.

The calculations similar to
those in Ref.~\cite{hp93} lead to the following expressions for the
$ee$ collision frequency and thermal conductivity,
\begin{equation}
  \nu_{ee}={36 n_e \alpha^2 \hbar^2 c I(u,\theta) \over \pi m_e^* k_{\rm B}T },
  \qquad
  \kappa_{ee}= {\pi^3 k_{\rm B}^3 T^2 \over 108 \alpha^2 \hbar^2 c I(u,\theta)}.
\label{result}
\end{equation}
Here, $\alpha=e^2/\hbar c$ is the fine structure constant, and
\begin{eqnarray}
   I(u,\theta) &=& { 1 \over u} \,
       \int_0^\infty {\rm d}w \, { w \, {\rm e}^w
       \over ({\rm e}^w-1)^2 } \, \int_0^1 {\rm d}x\,x^2\,(1-x^2)
\nonumber \\       
  && \times     \int_0^{ \pi} {{\rm d} \phi \over  \pi}\,(1-\cos \phi)
   \left| {1 \over 1 + (x \theta /w)^2\,\chi_l(x)} \right.
\nonumber \\   
  &&  - \left.
      { u^2 (1-x^2) \,\cos \phi \over
      1- u^2 x^2 + u^2 (x \theta /w )^2\,\chi_t(x)} \right|^2
\label{I}
\end{eqnarray}
is a dimensionless function of two variables,
\begin{equation}
   u \equiv v_e/c, \qquad \theta= \hbar v_e q_0/(k_{\rm B}T)
   = \sqrt{3} T_{pe}/T,
\label{betatheta}
\end{equation}
$T_{pe}=\hbar \omega_{pe}/k_{\rm B}$ being the electron
plasma temperature determined by the electron plasma frequency
$\omega_{pe}=\sqrt{4 \pi e^2 n_e/m^*_e}$.
Both components of the matrix element, $M^{(1)}_{fi}$
and $M^{(2)}_{fi}$,
give equal contributions into Eq.~(\ref{I}), and the interference
term is negligibly small because of
the small-momentum-transfer approximation.
Furthermore,
$w=\hbar \omega/k_{\rm B}T$;
$\phi$ is the angle between $\bm{p}_{1t}$ and
$\bm{p}_{2t}$, the components of
$\bm{p}_1$ and $\bm{p}_2$ transverse to $\bm{q}$;
and the integration
over $\phi$ is trivial. Equations (\ref{result}) and (\ref{I})
are natural generalizations of Eqs.\ (58) and (59) of Heiselberg and
Pethick \cite{hp93} to the case of a degenerate gas of
particles of arbitrary degree of relativity (in our case $v/c$
may be arbitrary while in Ref.\ \cite{hp93} $v=c$).
The difference of numerical factors in the
expressions for $\kappa$ in our Eq.\ (\ref{result}) and in Eq.\ (58)
of Ref.\ \cite{hp93} (108 versus 24)
stems from the difference of physical systems under consideration
(an electron gas versus a gas of light quarks interacting
through gluon exchange).

\section{Four regimes of electron-electron collisions}
\label{regimes}

Thus the collision frequency and the thermal conductivity
(\ref{result}) are solely determined by the electron
number density and the temperature. Their calculation
reduces to the calculation of the function $I(u,\theta)$
from Eq.\ (\ref{I}). Clearly, the function can be written
as
\begin{equation}
   I=I_l+I_t+I_{lt},
\label{Ilt}
\end{equation}
where $I_l$ is the contribution from the $ee$ interaction
via the exchange of
longitudinal plasmons (the first
term in the squared modulus);
$I_t$ comes from the interaction
via the exchange of transverse plasmons (the second
term); and $I_{lt}$ is the mixed
term.

The analysis reveals four regimes (I--IV) of $ee$
collisions in a strongly degenerate electron gas.
These regimes are summarized in Table \ref{tab:regimes}.
The regimes I and II are realized for non-relativistic
electrons, while in the regimes III and IV
electrons are ultrarelativistic.
The regimes I and III take place for sufficiently
high temperatures $T \gtrsim T_{pe}$, at which the Pauli
principle does not restrict energy transfers between
colliding electrons ($\hbar \omega < k_{\rm B}T$;
see, e.g., Lampe \cite{lampe68},
especially his Fig.\ 1).
The regimes II and IV refer to a colder electron gas,
where energy transfers are essentially limited by the Pauli
blocking (e.g., Lampe \cite{lampe68} and Flowers and
Itoh \cite{fi76}).

\begin{table}[b]
\caption[]{Four regimes of thermal
conduction of degenerate electrons owing to
$ee$ collisions.}
\label{tab:regimes}
\begin{center}
\begin{tabular}{c c c c c }
\hline
\hline
&~Electron~& &~~Main~~
& $T$-dependence \\
Regime&~velocity~&~Temperature~&~contribution~
& of $\kappa_{ee}$ \\
\hline 
I  & $v_e \ll c$ &  $T \gtrsim T_{pe}$ & $I_l$  &  $T^2/\ln(T/T_{pe})$
  \\
II  & $v_e \ll c$ &  $T \ll T_{pe}$ & $I_l$  & $1/T$
  \\
III  & $v_e \approx c$ &  $T \gtrsim T_{pe}$ & $I_l+I_t+I_{lt}$
&  $T^2/\ln(T/T_{pe})$
  \\
IV  & $v_e \approx c$ &  $T \ll T_{pe}$ & $I_t$ & const
  \\
\hline
\hline
\end{tabular}
\end{center}
\end{table}

The analysis of Eq.~(\ref{I}) gives the following asymptotic
values of $I$ in the different regimes.

In the regime I (where $u \lesssim 1$ and $\theta \lesssim 1$)
\begin{eqnarray}
  I_l&=&{1 \over u}\, \left( {2 \over 15} \ln {1 \over \theta }
       + 0.1657 \right),
\nonumber \\
  I_t&= & u^3 \,\left( {8 \over 315} \ln {1 \over \theta u }
       + 0.05067   \right),
\nonumber \\
  I_{lt}&=& u \,\left( {8 \over 105} \ln {1 \over \theta }
       + 0.1236 \right).
\label{regime1}
\end{eqnarray}
The logarithmic terms in brackets represent Coulomb
logarithms, while the second terms are the corrections
calculated using the standard technique
\cite{lampe68}. The leading contribution comes
from $I_l$. It was calculated by Lampe \cite{lampe68},
with a slightly less accurate correction term
($1.30 \times 2/15 \approx 0.173$ from his Eqs.~(5.22)
and (5.23), instead of our 0.1657).
Retaining this term in the regime I, one has $I=I_l$ and
\begin{equation}
    \kappa_{ee}={ 5 \pi^3 k_{\rm B}^3 T^2 v_e \over
    72 \alpha^2 \hbar^2 c^2  \, [\ln(1/\theta)+1.242] }.
\label{kappa1}
\end{equation}
Note that for $I_{t}$ the regime I extends to lower temperatures
$T \sim u T_{pe}$, than for $I_l$ and $I_{lt}$, but this
circumstance does not affect noticeably the thermal
conductivity $\kappa_{ee}$ because $I_t$ is relatively
insignificant in the given regime.

In the regime III (where $u \approx 1$ and $\theta \lesssim 1$),
\begin{eqnarray}
  I_l~& = & 2I_t={2 \over 15}\, \ln {1 \over \theta }
       + 0.1657,
\nonumber \\
  I_{lt}&=&{2 \over 15}\, \ln {1 \over \theta }
       + 0.1399,
\nonumber \\
   I~& = & {1 \over 3} \, \ln {1 \over \theta } + 0.3884.
\label{regime3}
\end{eqnarray}
Again, the logarithmic terms are Coulomb logarithms and the second
terms are corrections. In this case, all terms ($I_l$,
$I_t$, and $I_{lt}$) give comparable contributions into $I$.
The asymptote of $I$ in this regime was obtained
by Heiselberg and Pethick \cite{hp93} (their Eq.~(60),
with a slightly less accurate correction factor
0.30 instead of our 0.3884) in their studies of quark plasma.
For an electron plasma, the conductivity $\kappa_{ee}$
in the regime III was calculated by Urpin and Yakovlev
\cite{uy80}. Their result is equivalent to Eq.~(\ref{regime3})
for $I$ but with less accurate correction
($\ln(2)/3 \approx 0.231$ instead of 0.3884) because they
erroneously
used the approximation of static longitudinal electron screening
of the Coulomb interaction in all terms (with $\Pi_l=\Pi_t=q_0^2$ in
Eq.~(\ref{polariz})). Employing $I$ from Eq.~(\ref{regime3})
in the regime III we obtain
\begin{equation}
    \kappa_{ee}={ \pi^3 k_{\rm B}^3 T^2 \over
    36 \alpha^2 \hbar^2 c  \, [\ln(1/\theta)+1.165] }.
\label{kappa3}
\end{equation}

In the regimes II and IV (where $\theta \gtrsim 1$ and $u\leq 1$ is arbitrary)
we have
\begin{eqnarray}
  I_l={\pi^5  \over 15u \theta^3}, \quad
  I_t=2 \zeta(3)\,{ u \over \theta^2},\quad
 I_{lt}={\xi \over \theta^{8/3}}  \,u^{1/3}, &&
\label{regime24}\\
  \xi={(2\pi)^{2/3}\over 3} \,\Gamma\left(14 \over 3 \right)\,
  \zeta\left( 11 \over 3 \right)\approx 18.52,&&
\nonumber
\end{eqnarray}
where $\zeta(z)$ is the Riemann zeta function (with $\zeta(3)=1.202$)
and $\Gamma(z)$
is the gamma function.
The expression for $I_t$ (with $u=1$) was obtained by
Heiselberg and Pethick \cite{hp93} for an ultrarelativistic
quark plasma.

The asymptotic expressions (\ref{regime24}) are derived
from Eq.~(\ref{I}) with the screening functions
$\chi_l$ and $\chi_t$ given by Eq.~(\ref{polariz1}).
In this approximation the screening of the $ee$ interaction
via the exchange of longitudinal plasmons
(the first part of the matrix element in Eq.~(\ref{matelement}))
is described by $q^2+\Pi_l\approx q^2+ q_0^2$, which
is equivalent to the static Debye-type screening with the screening
wavenumber $q_{{\rm scr},l}=q_0$. The screening via the exchange
of transverse plasmons (the second part of the matrix element)
is more complicated. It is described by the denominator term
$q^2-\omega^2/c^2+\Pi_t \approx q^2 + iq_0^2 \pi \omega v_e /(4qc^2)$
that represents the dynamical screening via the Landau damping
of transverse plasmons, with $\hbar \omega \sim k_{\rm B}T \ll v_e \hbar q$
(i.e., with low phase velocities
$\omega/q \ll v_e$). In this case the effective screening
wavenumber $q_{{\rm scr},t}$  is evidently given by
$q^3_{{\rm scr},t} \sim q_0^2  \omega v_e
/c^2 \sim q_0^2 k_{\rm B}T v_e/(\hbar c^2)$.

Using Eqs.~(\ref{result}) and (\ref{Ilt}) we can decompose
the collision frequency $\nu_{ee}$ into the same parts as
$I$, $\nu_{ee}=\nu_{ee}^{(l)}+\nu_{ee}^{(t)}+\nu_{ee}^{(lt)}$.
One can easily see, that in the regimes II and IV the
partial frequency $\nu_{ee}^{(l)}$ can be estimated as
$\nu_{ee}^{(l)} \sim
n_e \alpha^2 (k_{\rm B}T)^2/(\hbar q_{{\rm scr},l}^3 \mu)$.
The collision frequency $\nu_{ee}^{(t)}$ can be estimated as
$\nu_{ee}^{(t)} \sim u^4 \,\nu_{ee}^{(l)}$ by replacing
$q_{{\rm scr},l} \to q_{{\rm scr},t}$.
The factor $u^4$ takes into account the reduction
of the $ee$ interaction via transverse plasmons in the nonrelativistic
electron gas.
The quantity $\nu_{ee}^{(lt)}$ can be estimated as
$\nu_{ee}^{(lt)} \sim u^2 \,\nu_{ee}^{(l)}$
with $q_{{\rm scr},l}^3 \to q_{{\rm scr},l}^2\, q_{{\rm scr},t}$.

In the regime II, where $\theta \gtrsim 1$ and $u \ll 1$,
the exchange of longitudinal plasmons dominates, and
we have $I \approx I_l$,
\begin{equation}
     \kappa_{ee}= { 5 \hbar q_0^3 v_e^4 \over 36 \pi^2 T \alpha^2 c^2}.
\label{kappa2}
\end{equation}
This leading part of $\kappa_{ee}$ in the regime II was calculated by
Lampe \cite{lampe68}. Note that the asymptotic expression
(\ref{regime24}) for $I_{t}$ at $u \ll 1$ is actually
valid not at $T \lesssim T_{pe}$, as the expressions for
$I_{l}$ and $I_{lt}$, but at lower $T \lesssim uT_{pe}$.
However, this circumstance is relatively unimportant because
it is $I_{l}$ which dominates in the regime II.

In the regime IV, where $\theta \gtrsim 1$ and $u \approx 1$,
the exchange of transverse plasmons dominates,
with $I \approx I_t$, and
\begin{equation}
     \kappa_{ee}= { \pi^3 k_{\rm B} c q_0^2 \over 216\,\zeta(3) \alpha^2}.
\label{kappa4}
\end{equation}
It is remarkable that in this regime $\kappa_{ee}$ becomes
temperature independent. This regime has been discussed in
detail by Heiselberg and Pethick \cite{hp93} for quark plasma.
For the electron plasma, it was considered by
Flowers and Itoh \cite{fi76} and also by Urpin and Yakovlev \cite{uy80}
but both groups erroneously used the approximation of
static longitudinal screening in all channels ($\chi_l=\chi_t=1$)
which strongly underestimates the efficiency of the plasma screening of the
$ee$ interaction in the ultrarelativistic electron gas.
If that approximation were true, one would
obtain $\nu_{ee}^{(l)}=\nu_{ee}^{(lt)}=
2 \nu_{ee}^{(t)}$ and $\nu_{ee}=2.5 \nu_{ee}^{(l)}$,
whereas in fact
$\nu_{ee}^{(l)} \ll \nu_{ee}^{(lt)} \ll \nu_{ee}^{(t)}$ in the regime IV,
and $\kappa_{ee}$ is significantly lower than predicted by the previous
calculations \cite{fi76,uy80}.

Let us note that the temperature behavior of $\kappa_{ee}$
(Table \ref{tab:regimes})
corresponds to an ordinary Fermi-liquid
(where $\kappa \propto 1/T$; e.g., Baym and Pethick \cite{bp91})
only in the regime II.
In the regimes I and III the plasma is too warm
(although degenerate) to reach the Fermi-liquid limit,
where energy transfers between colliding particles are
strongly restricted by the Pauli principle. In the regime IV
the plasma is cold but the matrix element of the $ee$
interaction essentially depends on energy transfers
$\hbar \omega$ owing to the Landau damping of transverse
plasmons which violates the Fermi-liquid
behavior \cite{hp93}.

To facilitate applications, we have calculated
the functions $I_l$, $I_t$ and $I_{lt}$ from
Eq.~(\ref{I}) for a dense grid of $u$ and
$\theta$ and fitted the numerical data by analytic
functions. We obtain
\begin{eqnarray}
  I_{l}&=& {1 \over u} \,  \left( 0.1587 -   
       {0.02538 
       \over 1+0.0435\, \theta} \right)      
\nonumber \\
     && \times  \ln \left( 1+ { 128.56 \over
      37.1\,\theta +10.83\, \theta^2 + \theta^3 } \right) .
\label{fitl}
\end{eqnarray}
Therefore, $I_l$ is inversely proportional to $u$,
with the proportionality coefficient dependent solely on $\theta$.
Furthermore,
\begin{eqnarray}
  I_{t}&=&  u^3 \, \left( {2.404 \over C}+
       {C_2-2.404/C \over 1+0.1\, \theta\,u} \right)
\nonumber \\
     && \times
      \ln \left( 1+ { C \over
      A \,\theta\,u + \theta^2\, u^2} \right) ,
\label{fitt}
\end{eqnarray}
where $A=20+450\,u^3$, $C=A \, \exp(C_1/C_2)$,
$C_1=0.05067+0.03216\,u^2$, and $C_2=0.0254+0.04127\,u^4$.
Finally,
\begin{eqnarray}
  I_{lt}&=&u \, \left( {18.52\, u^2 \over C}+
       {C_2-18.52\,u^2/C \over 1+0.1558\, \theta^B} \right) 
\nonumber \\
     && \times   \ln \left( 1+ { C \over A
      \theta +10.83\, \theta^2 u^2+ (\theta u)^{8/3}} \right) ,
\label{fitlt}
\end{eqnarray}
where $A=12.2+25.2\,u^3$, $B=1-0.75\,u$,
$C=A\, \exp(C_1/C_2)$,
$C_1=0.123636+0.016234\,u^2$, and
$C_2=0.0762+0.05714\,u^4$.
These fit expressions reproduce also all asymptotic limits
mentioned above. The maximum fit errors of $I_l$, $I_t$, and
$I_{lt}$ are, respectively, 0.2\% (at $u=0.61$ and $\theta=3.8$), 6.3\%
(at $u=0.878$ and $\theta \sim 0.175$, where
the electrons become nondegenerate)
and 8.4\% (at $u=0.19$ and $\theta=19$). The maximum fit
error of the total function $I$ in Eq.~(\ref{Ilt}) is 2.5\% (at $u=0.924$ and
$\theta=0.232$).

\section{Discussion}
\label{discussion}

\begin{figure}
    \begin{center}
    \leavevmode
    \epsfxsize=80mm
    \epsfbox[50 175 535 655]{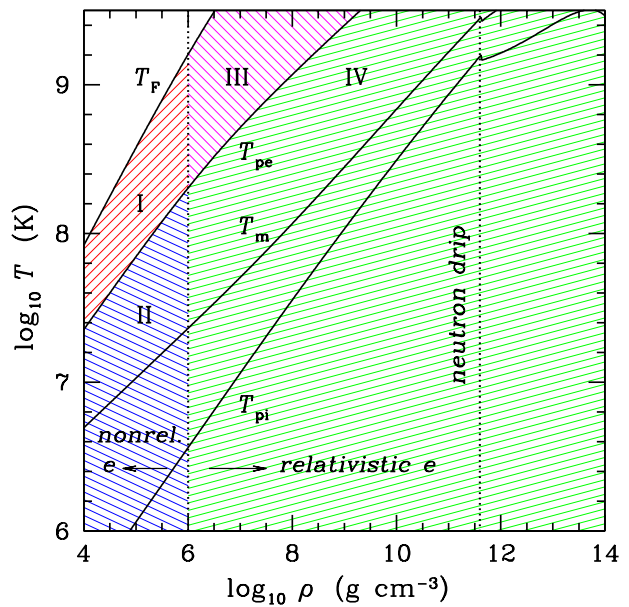}
    \end{center}
    \caption{(color online) Temperature-density diagram for cold-catalyzed
    matter. We show the electron degeneracy temperature
    $T_{\rm F}$, the electron plasma temperature $T_{pe}$,
    the melting temperature $T_{\rm m}$ of the ion crystal,
    and the ion plasma temperature $T_{pi}$. The left dotted
    line separates the regions of a low-density non-relativistic
    electron gas and a denser gas of ultrarelativistic
    electrons. The right dotted line indicates the neutron
    drip point. Shaded regions I--IV show the $T-\rho$
    domains of the different $ee$ collision regimes
    (Table \ref{tab:regimes}).
  }
\label{fig:phys}
\end{figure}

Let us discuss the efficiency of $ee$ collisions for
thermal conduction in a degenerate electron gas.
Figure \ref{fig:phys} shows the temperature-density
diagram for a dense matter. For certainty, we
have adopted the model of cold-catalyzed (ground-state) matter.
It is composed of electrons and ions (atomic nuclei)
at densities smaller than the neutron drip density
$\rho_{\rm nd} \approx 4 \times 10^{11}$ g~cm$^{-3}$
(the right vertical dotted line),
with an addition of free neutrons at higher densities
$\rho$. The composition of the cold-catalyzed matter
is taken from Haensel and Pichon \cite{hp94} at $\rho<\rho_{\rm nd}$
and from Negele and Vautherin \cite{nv73} at higher $\rho$.
Weak first-oder phase transitions which accompany
changes of ground-state nuclides with the growth of
$\rho$ are smoothed out as described by Kaminker et al.\
\cite{kaminkeretal99}. At $\rho \lesssim 10^8$ g~cm$^{-3}$
the ground-state matter is composed of iron ($^{56}$Fe).
The densities $\rho \lesssim 10^9$ g~cm$^{-3}$ are
appropriate for degenerate stellar cores of white
dwarfs and evolved stars; the densities $\rho \lesssim 10^{14}$
are appropriate for envelopes (crusts) of neutron stars.

In Figure \ref{fig:phys} we plot the electron degeneracy
temperature $T_{\rm F}=(\mu- m_ec^2)/k_{\rm B}$,
the electron plasma temperature
$T_{pe}$, the melting temperature of the ion crystal
$T_{\rm m}\approx Z_i^2 e^2/(175 a k_{\rm B})$,
and the ion plasma temperature
$T_{pi}=\hbar \omega_{pi}/k_{\rm B}$
(where $\omega_{pi}=\sqrt{4 \pi Z_i^2e^2 n_i/m_i}$ is
the ion plasma frequency,
$Z_i e$ is the ion charge, $m_i$ is its mass,
$n_i$ is the ion number density, and $a=(4 \pi n_i/3)^{-1/3}$
is the ion-sphere radius).
The left vertical dotted line separates the regions
of non-relativistic degenerate electrons ($v_e \ll c$,
$\rho \ll 10^6$ g~cm$^{-3}$) and ultrarelativistic
electrons ($v_e \to c$, $\rho \gg 10^6$ g~cm$^{-3}$).
The shaded regions I--IV show the
temperature-density domains with the different regimes of $ee$
collisions in a degenerate electron gas (see Section
\ref{regimes}, Table \ref{tab:regimes}).

In order to explore the importance of the $ee$ collisions
for thermal conduction let us compare $\kappa_{ee}$ with
the thermal conductivity $\kappa_{ei}$ owing to $ei$
collisions (Section \ref{introduc}). The conductivity
$\kappa_{ei}$ is calculated using the formalism
of Potekhin et al.\ \cite{pbhy99} and Gnedin et al.\ \cite{gyp01}.

\begin{figure}
    \begin{center}
    \leavevmode
    \epsfxsize=82mm
    \epsfbox{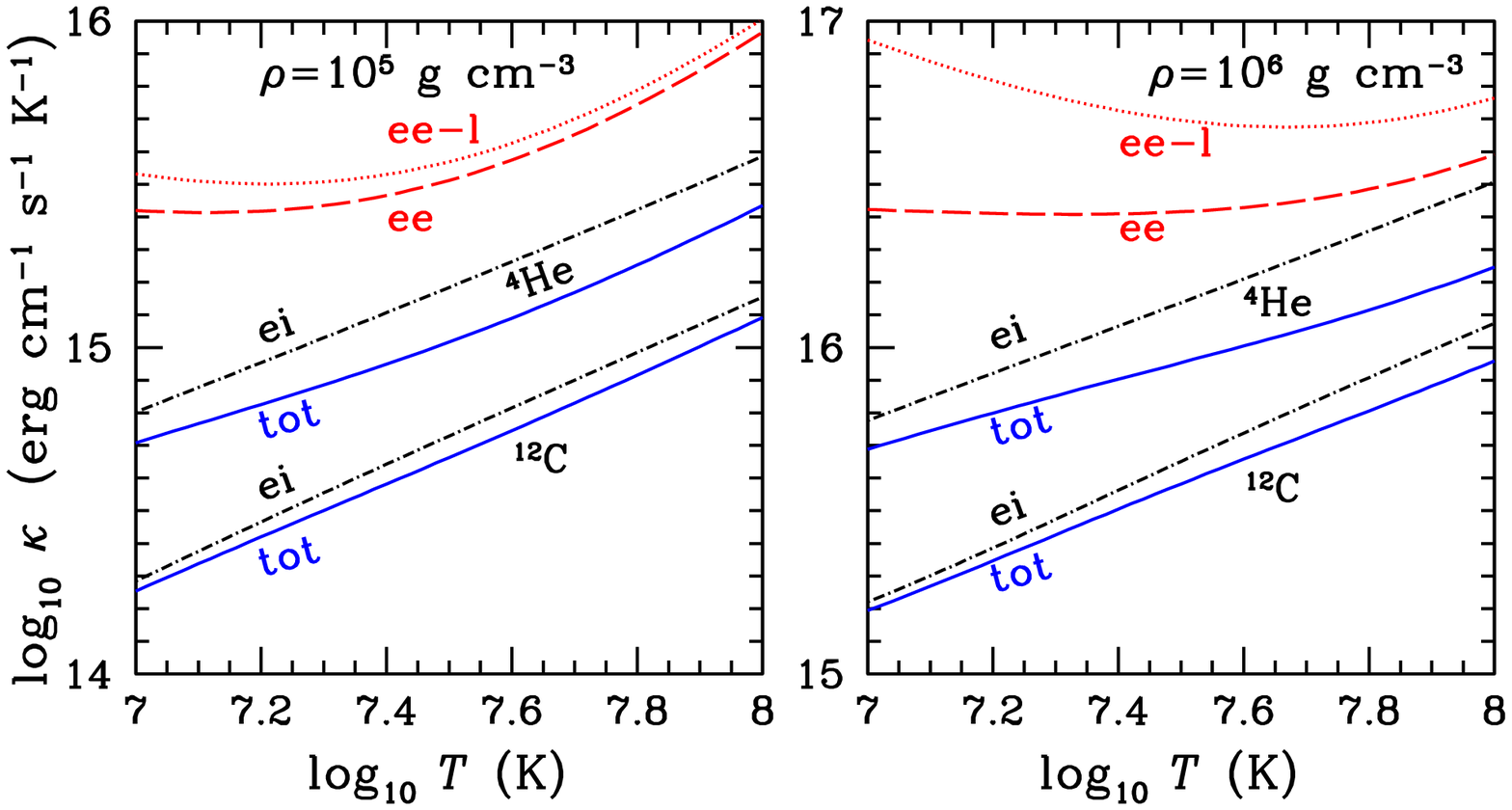}
    \end{center}
    \caption{(color online) Temperature dependence of the electron thermal
    conductivity at $\rho=10^5$ g~cm$^{-3}$ ({\it left})
    and $10^6$ g~cm$^{-3}$ ({\it right}). The dashed
    line $ee$ shows $\kappa_{ee}$; the dotted line $ee-l$
    is the same but retaining the contribution of
    longitudinal plasmons alone. The dot-and-dashed lines
    show $\kappa_{ei}$ and the solid lines show the
    total (tot, $ee+ei$) conductivity $\kappa_e$ for helium and carbon
    plasmas.
  }
\label{fig:rho56}
\end{figure}

In Figure \ref{fig:rho56} we plot the temperature dependence
of the electron thermal conductivity of helium ($^4$He)
or carbon ($^{12}$C) plasmas at $\rho=10^5$ g~cm$^{-3}$ ({\it left})
or $10^{6}$ g~cm$^{-3}$ ({\it right}). At $\rho=10^5$ g~cm$^{-3}$
the electrons are nonrelativistic while at
$\rho=10^6$ g~cm$^{-3}$ they are mildly relativistic.
According to Eq.~(\ref{kappae+i}), the total electron
thermal conductivity $\kappa_e$ is determined by the
partial contributions $\kappa_{ee}$ and $\kappa_{ei}$,
a minimum partial contribution being most important.
The total conductivities are shown by the solid lines;
$\kappa_{ei}$ by the dot-and-dashed lines;
$\kappa_{ee}$ is plotted by the dashed lines (the same
for the helium and carbon plasmas); the dotted lines give
$\kappa_{ee}$ neglecting the contribution of transverse
plasmons (i.e., by setting $I=I_l$).

For the densities and temperatures displayed in Figure \ref{fig:rho56},
the $ei$ collisions are mainly more efficient than the $ee$
ones, although the contribution of the $ee$ collisions
into the total thermal conductivity is noticeable.
One can see that the $ei$ collisions are more important
for heavier elements (because the Coulomb $ei$ scattering
cross section is much higher than the $ee$ scattering
cross section for high-$Z$ elements; see Lampe \cite{lampe68}).
The $ee$ collisions would be negligible for
the iron plasma if the data for this plasma
were displayed in Figure \ref{fig:rho56}. The $ee$ collisions are more
efficient in the helium plasma than in the carbon plasma
because the helium ions have lower charges.  For a given
chemical composition, $\kappa_{ee}$ gives highest contribution
into $\kappa_{e}$ at temperatures $T$ a few times lower than $T_{pe}$
(we have $\log_{10} T_{pe}{\rm [K]} \approx 7.86$ and
8.32 at $\rho = 10^5$ and $10^6$ g~cm$^{-3}$, respectively).
These temperatures separate the high-temperature
and low-temperature $ee$ collision regimes (e.g.,
the regimes I and II in the non-relativistic electron gas,
see Table \ref{tab:regimes}).
For $\rho=10^5$ g~cm$^{-3}$, the electron gas is only slightly relativistic.
Accordingly, the contribution of the Landau damping (transverse
plasmons) into $\kappa_{ee}$ is relatively
small, and the results of Lampe \cite{lampe68} are sufficiently
accurate. For higher $\rho=10^6$ g~cm$^{-3}$, the contribution
of the Landau damping becomes more important
which invalidates the previous results of Refs.\ \cite{fi76,uy80}.
In this case the Landau damping strongly reduces $\kappa_{ee}$
which makes the $ee$ collisions much more important for thermal
conduction than it was thought before. For still higher $\rho$ the
contribution of $\kappa_{ee}$ into $\kappa_{e}$ in a plasma of
light elements at $T \sim T_{pe}$ would be even more significant.
However, the temperature $T_{pe}$ would become so high that light
elements would start burning in thermonuclear reactions.
Therefore, $ee$ collisions can noticeably decrease the
electron thermal conductivity at densities and temperatures
important for nuclear burning of light elements (for instance,
in the vicinity of the carbon ignition curve). This should be
taken into account in simulations of nuclear evolution of
stars (e.g., carbon ignition in white dwarfs \cite{bhw04}).

\begin{figure}
    \begin{center}
    \leavevmode
    \epsfxsize=80mm
    \epsfbox[30 180 535 655]{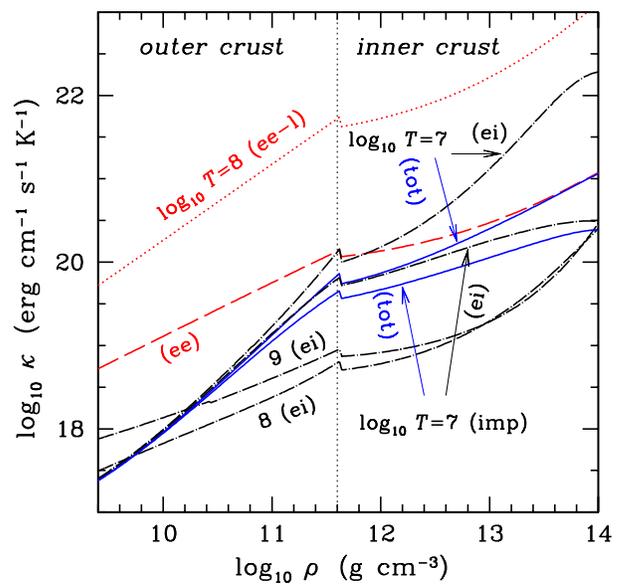}
    \end{center}
    \caption{(color online) Density dependence of the electron
    thermal conductivity at different $T$ (marked by
    the values of $\log_{10} T$ near the curves) in a neutron star
    crust composed of ground-state matter; the vertical dotted
    line is the neutron drip point.
    Dot-and-dashed lines ($ei$)
    show the conductivity $\kappa_{ei}$ owing to $ei$ collisions
    at $T=10^7$, $10^8$, and $10^9$~K in a crystal of atomic
    nuclei for pure ground-state matter.
    In addition, the lower dot-and-dashed line at $T=10^7$~K
    shows $\kappa_{ei}$ which includes the contribution
    of electron scattering by impurity nuclei at $T=10^7$~K
    (for $n_{\rm imp}/n_i=0.05$ and $|Z_{\rm imp}-Z_i|=2$).
    The dashed line is $\kappa_{ee}$; it
    is almost independent of $T$. The dotted
    line is the partial conductivity $\kappa_{ee}^{(l)}$
    mediated by longitudinal plasmons alone at $T=10^8$~K; it scales
    as $T^{-1}$. The solid lines show the total electron
    conductivity $\kappa_e$ at $T=10^7$ K for the matter without
    and with impurities. If $T \gtrsim 10^8$~K the effect
    of impurities is weak and $\kappa_e \approx \kappa_{ei}$.
    }
\label{fig:econd}
\end{figure}

Figure \ref{fig:econd} shows the density dependence of the
electron thermal conductivity
in the density range from $\sim 2.5 \times 10^9$
g~cm$^{-3}$ to $10^{14}$ g~cm$^{-3}$ for the three values of
the temperature $T=10^7$, $10^8$, and $10^9$~K.
We employ the same model of ground-state
matter as in Figure \ref{fig:phys}.
The displayed density range is appropriate to a crust
of a neutron star. The vertical dotted line is
the neutron drip point which separates the crust into
the outer and inner parts.
The adopted values of $T$ and $\rho$ refer to the regime
IV of the $ee$ collisions, where the electron gas is
ultrarelativistic, $T \ll T_{pe}$, and the Landau damping
dominates. The conductivity $\kappa_{ee}$ (the dashed line)
is well approximated by the temperature
independent conductivity (\ref{kappa4})
governed by the Landau damping; $\kappa_{ee}$
slightly deviates from Eq.\ (\ref{kappa4}) only at lowest $\rho$
for $T=10^9$ in Figure \ref{fig:econd}, where $T$ is still
insufficiently lower than $T_{pe}$
(see Figure \ref{fig:phys}). Retaining the longitudinal contribution
into $\kappa_{ee}$ ($I=I_l$) and setting $T=10^8$ K, we would get
the dotted curve in Fig.\ \ref{fig:econd}. It goes much
higher than the dashed curve indicating that this longitudinal
contribution is really insignificant.

For comparison, the dot-and-dashed lines
in Figure \ref{fig:econd} give the thermal conductivity
$\kappa_{ei}$ (mostly from
Figure 4 of Ref.\ \cite{gyp01})
for the same values of $T$, and the solid lines
give the total conductivity $\kappa_e$.
For $T=10^8$ and $10^9$, the effects of possible impurities
in dense matter
(atomic nuclei with charge numbers $Z_{\rm imp}$ different
from charge numbers $Z_i$ of ground-state nuclei)
are expected to be small (e.g., Gnedin et al.\ \cite{gyp01}).
At $T=10^7$~K, the effects of impurities can be substantial.
For this $T$ in Figure \ref{fig:econd}
we present $\kappa_{ei}$ and $\kappa_e$ for pure ground-state
matter and also for the matter
which contains impurities with $|Z_{\rm imp}-Z_i|=2$ and
with the fractional number of impurity nuclei of $n_{\rm imp}/n_i=0.05$.
The impurities increase the $ei$ collision rate and
decrease the conductivities $\kappa_{ei}$ and $\kappa_e$.

As seen from Figure \ref{fig:econd}, $\kappa_{ei}$ is
more important than $\kappa_{ee}$ at all displayed densities
for $T=10^8$ and $10^9$~K. In these cases $\kappa_e \approx \kappa_{ei}$
and we do not show $\kappa_e$ to simplify the figure.
However, $\kappa_{ee}$ dominates in the pure ground-state matter
for $T=10^7$~K
at $\rho \gtrsim 10^{12}$ g~cm$^{-3}$ 
(in the inner crust of a neutron star).
This dominance is fully produced by the Landau damping
of transverse plasmons in the $ee$-collisions.
For $\rho \gtrsim 10^{13}$ g~cm$^{-3}$ in this case we
have $\kappa_e \approx \kappa_{ee}$.
For an impure cold-catalyzed matter $\kappa_{ei}$
remains important for all densities at $T=10^7$~K,
but $\kappa_{ee}$ is comparable to $\kappa_{ei}$
in the inner neutron star crust.
Note that the values of $\kappa_{ei}$ in Figure \ref{fig:econd}
take into account, in an approximate manner, the freezing
of Umklapp processes in electron-phonon scattering
at low temperatures (see, e.g., Ref.\ \cite{gyp01}).
For $T \gtrsim 10^8$~K the freezing
is unimportant, but at $T \sim 10^7$~K it enhances $\kappa_{ei}$.
A more rigorous treatment of this freezing can partly
remove this enhancement (A.~I. Chugunov,
private communication), which can somewhat decrease $\kappa_{ei}$
and reduce the importance of $\kappa_{ee}$.

Therefore, a correct treatment of
$ee$ collisions can considerably reduce the
electron thermal conductivity in a cold
neutron star crust, with $T \sim 10^7$~K.
This can increase the time of heat diffusion
from the inner crust to the surface \cite{ur01}. In
particular, the effect can be important for the propagation of thermal
waves produced by pulsar glitches  and for the
emergence of these waves on the pulsar surface.
The emergence can be, in principle, observed and
give valuable information on the nature of pulsar glitches \cite{ll02}.  

While calculating $\kappa_{ee}$ we have taken into
account only the electron contribution
into the polarization functions $\Pi_l$ and
$\Pi_t$ in Eq.\ (\ref{polariz}) and
neglected the ion contribution.
As shown by Lampe \cite{lampe68} this is
a good approximation at high temperatures
$T \gtrsim Z^2e^2/(a k_{\rm B})$ at which the ions
are weakly coupled and constitute a Boltzmann gas.
The calculation of the ion polarization functions at lower $T$
(in the regime of strong ion coupling, in an ion liquid
or solid) is a complicated problem whose detailed
solution is still absent. We have also neglected
the effects of strong magnetic fields which can
greatly modify electron heat transport in
strongly magnetized neutron star envelopes.
These effects are numerous (e.g., Yakovlev and Kaminker
\cite{yk94}) and are subdivided
into classical (owing to rapid electron Larmor rotation)
and quantum ones (owing to the Landau quantization of
electron motion in a magnetic field).
In principle, a proper treatment of $ee$ collisions
in a strong magnetic field can be performed
in the same framework of the dynamical plasma screening
as used above. However, the problem
becomes much more complicated because
the electron polarization tensor in a magnetic field
is anisotropic (depends on the relative
orientations of the wavevector $\bm{q}$ and the magnetic field)
and cannot be generally
decomposed into purely longitudinal and transverse
parts \cite{abr84}.
The effects of ion polarization and strong magnetic fields
are outside the scope of the present paper.

\section{Conclusions}
\label{conclusions}

We have reconsidered the electron thermal conductivity
$\kappa_{ee}$ of degenerate electrons
produced owing to the $ee$ collisions
taking into account the Landau damping due to the exchange
of transverse plasmons (following the calculations
of kinetic properties of quark plasma by Heiselberg and
Pethick \cite{hp93}). The Landau damping has been neglected
in all previous calculations of $\kappa_{ee}$.
We have analyzed the four regimes of
the $ee$ collisions in the degenerate electron gas
(Section \ref{regimes}, Table \ref{tab:regimes}) and
obtained analytic expressions for $\kappa_{ee}$ which
accurately approximate the results of numerical calculations
of $\kappa_{ee}$ in wide ranges of
the temperature and density of the matter. These
results can be applied to study thermal conduction
of degenerate electrons in metals, semiconductors,
in degenerate cores of evolved stars
and white dwarfs, and in envelopes of neutron stars.

Our main conclusions are the following.

(1) The Landau damping strongly modifies
$\kappa_{ee}$ in a relativistic degenerate electron
gas, at densities $\rho \gtrsim 10^6$~g~cm$^{-3}$,
but it is also quite noticeable at lower $\rho$
(for instance, at $\rho=10^5$~g~cm$^{-3}$, see Figure
\ref{fig:rho56}). The Landau damping increases the
$ee$ collision rate and decreases $\kappa_{ee}$,
increasing the contribution of the $ee$ collisions into the total
electron thermal conductivity $\kappa_{e}$.

(2) The most dramatic effect of the Landau damping
on $\kappa_{ee}$ takes place at $\rho \gg 10^6$~g~cm$^{-3}$
for temperatures $T$ much below the electron plasma temperature
$T_{pe}$ (the regime IV, Table \ref{tab:regimes}).
In this case $\kappa_{ee}$ shows a non-Fermi-liquid
behavior; it becomes temperature independent and is
described by the asymptotic expression (\ref{kappa4}).

(3) The conductivity $\kappa_{ee}$ becomes comparable
to the electron thermal conductivity $\kappa_{ei}$ provided by the
electron-ion collisions (and gives thus a noticeable contribution
into the total conductivity $\kappa_e$) in a warm plasma of light
(low-$Z$) ions at temperatures $T$ a few times lower than $T_{pe}$
(Figure \ref{fig:rho56}). These conditions are typical for
degenerate cores of white dwarfs and giant stars where
thermonuclear burning of light elements can occur
(particularly, in the vicinity of the carbon ignition curve).

(4) The conductivity $\kappa_{ee}$ dominates over $\kappa_{ei}$
and determines the total electron thermal conductivity $\kappa_e$
of the dense pure cold-catalyzed matter at $T \sim 10^7$ K
and $\rho \gtrsim 10^{12}$~g~cm$^{-3}$ (in the inner crust
of a cold neutron star, Figure \ref{fig:econd}).
At these $T$ and $\rho$ the conductivity
$\kappa_{ee}$ is important even for
impure cold-catalyzed matter. It can
affect the propagation of thermal waves,
excited in the inner neutron star crust during pulsar glitches, 
to the pulsar surface.

The electron conductivity $\kappa_{e}$ operates also in neutron star
cores, at $\rho \gtrsim 1.5 \times 10^{14}$ g~cm$^{-3}$.
This conductivity should also be reconsidered taking into account the
Landau damping. Our present results cannot be directly
applied to this case because in the cores
the electrons collide efficiently at least with degenerate
electrons, muons, and protons, and these collisions
deserve a special study.
We intend to analyze them in a subsequent publication.

\begin{acknowledgments}
We are grateful to A.~I.\ Chugunov 
and A.~Y.\ Potekhin for useful discussions.
This work was partly supported by
the Dynasty Foundation,
by the Russian Foundation for Basic Research
(grants 05-02-16245, 05-02-22003), and
by the Federal Agency for Science and Innovations
(grant NSh 9879.2006.2).
\end{acknowledgments}

\end{document}